\documentclass[aip,jcp,reprint]{revtex4-1}
\usepackage{amsmath}
\usepackage{graphicx}
\begin{document}

\title{
Clogging and Transport of Driven Particles in Asymmetric Funnel Arrays
} 
\author{C. J. O. Reichhardt}
\altaffiliation{Corresponding author email: cjrx@lanl.gov}
\affiliation{Theoretical Division and Center for Nonlinear Studies,
Los Alamos National Laboratory, Los Alamos, New Mexico 87545, USA}

\author{C. Reichhardt}
\affiliation{Theoretical Division and Center for Nonlinear Studies,
Los Alamos National Laboratory, Los Alamos, New Mexico 87545, USA}

\begin{abstract}
We numerically examine the flow and clogging of particles driven through asymmetric funnel arrays when the commensurability ratio of the number of particles per plaquette is varied.  The particle-particle interactions are modeled with a soft repulsive potential that could represent vortex flow in type-II superconductors or driven charged colloids.  The velocity-force curves for driving in the easy flow direction of the funnels exhibit a single depinning threshold; however, for driving in the hard flow direction, we find that there can be both negative mobility where the velocity decreases with increasing driving force as well as a reentrant pinning effect in which the particles flow at low drives but become pinned at intermediate drives. This reentrant pinning is associated with a transition from smooth one-dimensional flow at low drives to a clogged state at higher drives that occurs when the particles cluster in a small number of plaquettes and block the flow.  When the drive is further increased, particle rearrangements occur that cause the clog to break apart.  We map out the regimes in which the pinned, flowing, and clogged states appear as a function of plaquette filling and drive.  The clogged states remain robust at finite temperatures but develop intermittent bursts of flow in which a clog temporarily breaks apart but quickly reforms.
\end{abstract}

\maketitle

A wide variety of both hard and soft matter systems 
can be modeled as particles that exhibit depinning phenomena under 
an external drive \cite{1}. 
Specific examples include vortices in type-II superconductors \cite{2,3,4},
sliding charge density waves \cite{5},
the depinning of electron crystals \cite{6,7} or colloidal assemblies \cite{8,9,10}, 
and skyrmions in chiral magnets \cite{11,12}. 
In such systems,
the depinning and subsequent sliding dynamics can be characterized using features
in the velocity-force ($v-F$) curves.
The particle velocity is zero in the pinned phase
and becomes finite at a critical drive $F_{c}$.
Within the moving phase
the
velocity can increase linearly with the driving force
or it can show a series of 
steps corresponding to different effective depinning thresholds \cite{1}. 
It is also possible to observe negative differential mobility in which
the velocity drops with increasing
driving force due to changes in the flow pattern
or the onset of different modes of dissipation
\cite{1,13,14,15,16}.  In some 
cases, the particles move in the direction opposite to
the applied drive, producing
absolute negative mobility
\cite{17,18}.
Many studies involve particles
moving over an effectively
two-dimensional (2D) substrate;
however, depinning phenomena and nonlinear transport can also arise for particle 
flow in confined geometries such as quasi-one dimensional (1D)
channels,
constrictions 
\cite{19,20,21,22,23}, bottlenecks \cite{24,25,26,27}, 
or asymmetric funnel arrays \cite{28,29,30,31,32,33}.
Pinning in these systems
is produced by a confining effect of the walls, and 
particles that get stuck along the walls can
reduce or stop the flow of other particles due to the
particle-particle interactions.
This pinning mechanism generates
strong commensuration effects in which the
critical drive $F_c$ is enhanced
when the number of particles is
an integer multiple of the number
of plaquettes or of any other periodicity in the system.
Confined flow also appears in 
the motion of particulate matter through confining regions
such as hoppers and bottlenecks, 
where clogging effects occur due to
steric particle-particle interactions
\cite{34,35,36,37,38,39,40,41}.
Similar effects arise in the flow of
pedestrians or active matter
through constrictions \cite{42,43,44}.  

In this work we consider an assembly of particles driven through a linear array of
asymmetric funnels.  The soft repulsive interactions
between particles are modeled as a Bessel function $K_{1}(r)$ which 
decays exponentially at large $r$.
This specific interaction
potential describes vortices in type-II superconductors;
however, a variety of simulation studies have shown
that many of the behaviors observed for vortices interacting
with pinning landscapes
are generic to other systems of particles with similar
interactions,
such as charge-stabilized colloids with Yukawa interactions 
or magnetic colloids with dipolar interactions \cite{1,4,7,14,45,46}.
Previous work on particles driven through funnel arrays
focused on motion in the easy flow direction where the
depinning force is low,
and showed that a clear commensurability effect occurs
whenever the number of particles is an integer multiple of the number
of funnel plaquettes, 
while there is an overall increase in the depinning force with 
increasing commensurability ratio since crowding effects
make 
it more difficult to force particles through the
funnels \cite{30}. Here we consider
particles driven in the hard direction of the funnel asymmetry.
We find distinctive $v-F$ relations, 
include a clogging effect associated with reentrant pinning for
increasing drive.
In most systems that exhibit depinning phenomena, the velocity 
increases with increasing drive;
however, we find that there are extended regions 
of drive and filling ratios in which reentrant pinning occurs.
An initial depinning occurs at a low driving threshold $F_{c1}$ when
the particle density is uniform throughout the system, but at higher
drives clogging events occur in which
particles pile up in one or more of the funnels and block the flow,
so that the velocity drops back to zero.
As the drive increases further,
particles can eventually force their way through
the clog and allow the system to flow again above a second
depinning threshold $F_{c2}$.

We argue that our results are consistent with clogging rather than
jamming behavior.
In a jammed state,
particle-particle interactions
cause the system to behave like a uniform rigid solid
above a unique jamming density \cite{47}.
In our system, although the initially pinned state and the
flowing states are generally uniform,
the reentrant pinned state is
heterogeneous with 
local regions of high particle density.
The reentrant pinning
can occur for a wide range of fillings, as has been observed
for clogging transitions \cite{48},
rather than only at a specific filling, as
in the jamming transition.
We discuss how our results could be relevant to vortices in superconductors, 
colloids in constriction geometries, Wigner crystals,
and skyrmion systems,
as well as how the behavior we observe is different from that found for
particles
with hard disk interactions such as grains or bubbles.  

\begin{figure}
  \includegraphics[width=0.48\textwidth]{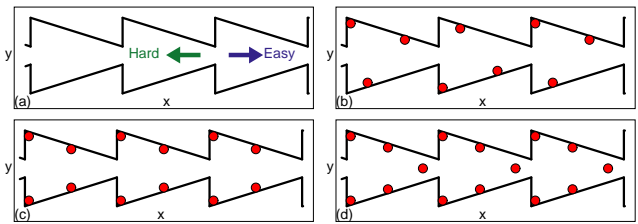}
  \caption{(a) Illustration of a portion of the system showing the
    asymmetric funnel walls.
    In the initial state, each funnel
    plaquette holds $N_c$ particles.  A drive $F_D$ is applied in
    the easy (right blue arrow) or hard (left green arrow) direction
    along the $x$ axis.
    The full sample contains $N_{pl}=16$ funnel plaquettes.
    (b,c,d) Illustrations of representative $F_D=0$
    particle configurations at
    $N_{c} =$ (b) 3.0, (c) $4.0$, and (d) $5.0$.
 }
\label{fig:1}
\end{figure}

{\it Simulation --}
We model a two-dimensional system
with periodic boundaries in the $x$ direction containing walls that form
an asymmetric funnel array, as illustrated
in Fig.~\ref{fig:1}.
There are a total of $N_{pl}=16$ funnel plaquettes,
each of which initially contains $N_c$ particles so that the
total number of particles is $N_p=N_cN_{pl}$.
The dynamics of particle $i$
is governed by the following overdamped equation of motion:
\begin{equation}
  \eta \frac{d{\bf R}_{i}}{dt} = -\sum_{ i \neq j}^{N_{p}}\nabla U_{cc}(R_{ij})
  - {\bf F}^{i}_{\rm wall} + {\bf F}_{D} + {\bf F}^{T}_{i} \ .
\end{equation}
We set the damping constant $\eta = 1.0$.
The particle-particle interactions are repulsive and have the
form
$U_{cc}(R_{ij}) = A_{0}K_{0}(R_{ij}/\lambda)$, where
$R_{ij}=|{\bf R}_i-{\bf R}_j|$,
${\bf R}_{i(j)}$ is the position of particle $i(j)$,
and $K_0$ is a Bessel function. 
We take $\lambda$ to be the unit of distance in our simulation.
The particle-particle interaction force
is given by $K_{1}$, which decays exponentially for $R_{ij} > \lambda$,
allowing us to cut off the interactions for
$R_{ij} > 6\lambda$. 
The unit of force in our simulation is $A_{0}$.
This model was previously used to study static
vortex configurations in a funnel geometry
as well as depinning and dynamics
for driving in the easy flow direction \cite{30}. 

The confining walls of each funnel plaquette are
composed of 4 repulsive elongated potential barriers so that
the entire system contains $N_b=4N_{pl}$ barriers.
Each barrier consists of a rectangular repulsive region with
a half-parabolic repulsive cap on each end.
Particle-barrier interactions are described by
${\bf F}^{i}_{\rm wall} = (f_{p}/r_{p})
\sum^{N_{b}}_{k=1}[
R_{ik}^{\pm}\Theta(r_{p}  - R_{ik}^{\pm})\Theta(R_{ik}^{||} -l_{k}){\hat {\bf R}}_{ik}^{\pm} + 
R_{ik}^{\perp}\Theta(r_{p}-R_{ik}^{\perp})\Theta(l_{k} - R_{ik}^{||}){\hat {\bf R}}_{ik}^{\perp}]$,
where
$f_p=15A_0$,
$R_{ik}^{\pm} = |{\bf R}_{i} - {\bf R}^{p}_{k} \pm l_{k}{\hat {\bf p}}^k_{||}|$,
$R_{ik}^{\perp,||} = |({\bf R}_{i} - {\bf R}^{p}_{k}) \cdot {\hat {\bf p}}^k_{\perp,||}|$,
${\bf R}^{p}_{k}$ is the position of the center point of barrier $k$,
and
${\hat {\bf p}}^k_{||}$ and ${\hat {\bf p}}^k_{\perp}$ are unit vectors
parallel and perpendicular, respectively, to the axis of barrier $k$.
The central rectangular barrier
is of size $2l_k=2.8\lambda$ for the vertical walls and
$2l_k=18\lambda/\sqrt{3}$ for the slanted walls.
The barriers are arranged to form mirror-symmetric sawtooth shapes that
produce the funnel array
illustrated in Fig.~\ref{fig:1}(a).

Initially $N_c$ particles are placed inside each funnel plaquette
and allowed to thermally relax under simulated annealing using
Langevin kicks ${\bf F}^{T}$ that obey
$\langle {\bf F}_i^{T}(t)\rangle = 0$
and $\langle {\bf F}^{T}_{i}(t){\bf F}^{T}_{j}(t^{\prime})\rangle
= 2\eta k_{B}T\delta_{ij}\delta(t - t^{\prime})$,
where $k_{B}$ is the Boltzmann constant.
Once the temperature has been reduced to zero,
we apply a driving force ${\bf F}_{D} = CF_{D}{\hat {\bf x}}$
with $C=+1$ for easy direction driving and $C=-1$ for hard
direction driving.
For each value of $F_D$ 
we measure the average particle velocity
$\langle V\rangle = N_{p}^{-1}\sum^{N_{p}}_{i}{\bf v} _{i}\cdot {\hat {\bf x}}$.
We first consider driving in the absence of thermal fluctuations, but
later we discuss
the effect of finite temperature on the clogging dynamics. 

\begin{figure}
  \includegraphics[width=0.48\textwidth]{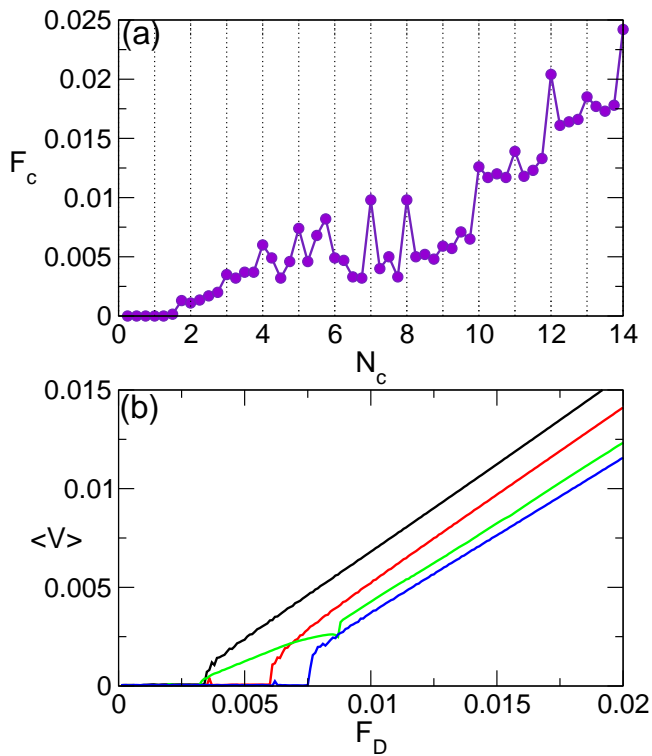}
  \caption{(a) The depinning force $F_{c}$ vs
    filling fraction $N_{c}$ for driving in the easy direction
    with $C=+1$.
    (b)
    Representative
    $\langle V\rangle$ vs
    $F_{D}$ curves for driving in the easy direction
    at $N_{c} = 3.0$ (black),
    $4.0$ (red), $4.5$ (green) and $5.0$ (blue).
    In each case there is a single depinning threshold $F_{c}$.  
}
\label{fig:2}
\end{figure}

{\it Results --}
Illustrations of representative
$F_{D} = 0$ particle configurations
obtained through the annealing procedure appear
in Fig.~\ref{fig:1}(b,c,d) for
$N_{c} = 3.0$, 4.0, and $5.0$, respectively.
At integer values of $N_c$, each plaquette captures $N_c$ particles,
while for noninteger values such as $N_c=5.5$, there is a mixture of
plaquettes containing 5 or 6 particles.
A more complete study of the particle ordering at the
zero drive condition is provided in Ref.~\cite{30}.
When a drive is applied in the positive $x$ or easy flow direction,
as in Ref.~\cite{30}, the depinning force depends strongly on
the commensuration ratio,
as shown in Fig.~\ref{fig:2}(a).
We plot four representative $\langle V\rangle$ versus $F_D$ curves
in Fig.~\ref{fig:2}(b)
for driving in the easy flow direction
at $N_{c} = 3.0$, 4.0, 4.5, and $5.0$.
In each case, there is a single
depinning threshold and $\langle V\rangle$ increases
monotonically with increasing $F_D$.
Other fillings produce similar $v-F$ curves.

\begin{figure}
\includegraphics[width=0.48\textwidth]{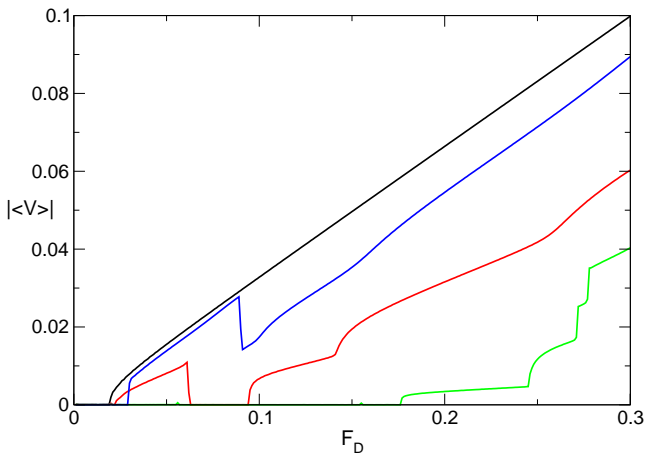}
\caption{ 
  The average particle velocity
  $|\langle V\rangle|$ vs $F_{D}$
  for driving in the hard direction with $C=-1$ for $N_{c} =  3.0$ (black),
  $3.25$ (red), $3.5$ (green), and $3.75$ (blue).
  There is a single depinning transition for $N_{c} = 3.0$ and
  a region of reentrant pinning where $|\langle V\rangle|$
  drops back to zero for $N_{c} = 3.25$.  An example
  of negative differential conductivity appears for $N_{c} = 3.75$
  in the form of a drop in
  $|\langle V\rangle|$ to a finite value near $F_{D} = 0.09$.  
}
\label{fig:3}
\end{figure}

\begin{figure}
\includegraphics[width=0.48\textwidth]{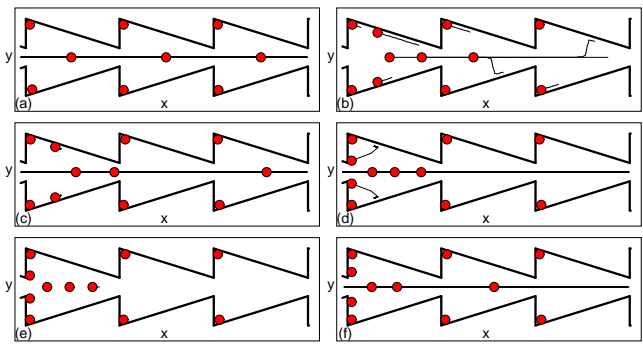}
\caption{ 
  The particle locations (circles) and trajectories (lines) for
  driving in the
  hard direction for the system in Fig.~\ref{fig:3}.
  (a) $N_{c} = 3.0$ at $F_{D} = 0.11$.
  (b-e) $N_c=3.25$ for different $F_D$.
  (b) $F_D=F_{c1} = 0.022$.
  (b) The
  first sliding phase at $F_{D} = 0.044$.
  (c) The onset of the reentrant pinned phase
  at $F_D=F_{rp} = 0.0625$.
  (d) The reentrant pinned or
  clogged phase at $F_{d} = 0.088$, where there is an accumulation of particles
  in one of the plaquettes that blocks the flow of the other particles.
  (e) The sliding state at $F_D=0.11$ above $F_{c2}$
  where the particles can move through the clog.       
}
\label{fig:4}
\end{figure}

In Fig.~\ref{fig:3} we plot
$|\langle V\rangle|$ versus $F_{D}$
for driving in the negative $x$ or hard direction
for $N_{c}=  3.0$, $3.25$, $3.5$, and $3.75$.
Here there are a variety of distinct dynamic behaviors and $v-F$ curve
characteristics that are very different from those found for driving in the
easy direction.
For the commensurate filling of $N_{c} = 3.0$, there 
is a single depinning transition
at $F_{c} = 0.02$ followed by a linear increase in
$|\langle V\rangle|$ with increasing $F_D$.
In Fig.~\ref{fig:4}(a) we show the particle trajectories 
for $N_{c}= 3.0$ at $F_{D}= 0.11$
where two particles are trapped at the corners 
of each plaquette and the remaining third of the 
particles move in a 1D channel
along the negative $x$-direction.  

In Fig.~\ref{fig:3} at $N_{c} = 3.75$, there is still a single     
depinning threshold at $F_{c} = 0.0275$;
however, a drop in $|\langle V\rangle|$
occurs near $F_{D} = 0.9$
indicating negative differential
mobility  with $d\langle V\rangle /dF_{D} < 0$.
For $N_{c} = 3.25$ we find a multi-step depinning process
with an initial depinning at
$F_{c1} = 0.022$.  Above $F_{c1}$,
$|\langle V\rangle |$ increases with
increasing $F_{D}$ until a drop to
$|\langle V\rangle| = 0$ occurs
over the interval of $0.0625 < F_{D} < 0.094$.
This reentrant pinned phase
is followed by a second 
depinning transition at $F_{c2} = 0.094$.
In Fig.~\ref{fig:4}(b) we show the particle positions and
trajectories
just at $F_{c1}$, while Fig.~\ref{fig:4}(c) illustrates
the particle flow in the first flowing region at
$F_{D} = 0.044$.  The transition into the reentrant
pinned state appears in Fig.~\ref{fig:4}(d) for $F_D=0.0625$.
The reentrant pinning is triggered when
two of the trapped particles in the plaquette shift
their positions from sitting next to a slanted wall of the funnel to
sitting next to a vertical wall of the funnel close to the leftmost
funnel opening, as shown in Fig.~\ref{fig:4}(d).
The shifted particles create
a stronger repulsive barrier for the passage of other particles through
the constriction on the left end of the funnel plaquette, causing the
previously moving particles to be blocked,
as plotted in Fig.~\ref{fig:4}(e)
for the reentrant pinned phase at $F_{D} = 0.088$. 
The clog is characterized by the accumulation of an excess
number of particles in a single plaquette.
In this case, a similar clogging configuration appears
in roughly every fourth plaquette,
so the reentrant  
pinning can be viewed as a clogging event dominated by
a small number of plaquettes.
As $F_{D}$ is
further increased, there
is a second depinning transition
at $F_{c2}$ to the second moving state
illustrated in Fig.~\ref{fig:4}(f) at $F_{D} = 0.11$.
Additional upward jumps in $|\langle V\rangle|$ can occur
at higher $F_D$ when 
additional particles 
in the clogged region become dislodged and join the flow.

\begin{figure}
\includegraphics[width=0.48\textwidth]{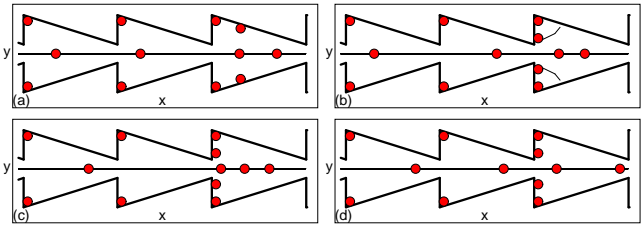}
\caption{
  The particle locations (circles) and trajectories (lines) for
  driving in the
  hard direction for the system in Fig.~\ref{fig:3} 
  at $N_{c} = 3.75$ where there is a region of
  negative differential conductivity. 
  (a) The flowing phase at $F_{D} = 0.066$.
  (b) The shift in particle positions from the slanted walls to the
  funnel aperture at the drop in $|\langle V\rangle |$ at $F_D=0.09$.
  (c) At $F_{D} = 0.12$, above the drop in $|\langle V\rangle |$,
  a partial clog forms when the particles next to the funnel aperture
  slow down the flow.
  (d) At $F_{D} = 0.26$, the size of the partial
  clog decreases.
}
\label{fig:5}
\end{figure}

A clogging phenomenon is also associated with
the jump down in
$|\langle V\rangle|$ versus $F_D$ for $N_{c} = 3.75$,
but in this case the clogging is only 
partial.
In Fig.~\ref{fig:5}(a) we
plot the 1D channel flow pattern in this system at $F_{D} = 0.066$ 
prior to the drop in $|\langle V\rangle |$.
As $F_D$ increases, some trapped particles suddenly shift from
sitting along the slanted walls to sitting next to the funnel
aperture, as shown
in the rightmost plaquette of Fig.5(b) at
$F_{D} = 0.09$.
These particles constrict but do not stop the channel flow, resulting in
a local accumulation of particles at the constriction, as illustrated
in Fig.~\ref{fig:5}(c) at $F_{D} = 0.012$.
As $F_{D}$ is further increased,
the flow gradually increases and the size of
the partial clog reduces, as shown  
in Fig.~\ref{fig:5}(d) at $F_{D} = 0.26$.

\begin{figure}
\includegraphics[width=0.48\textwidth]{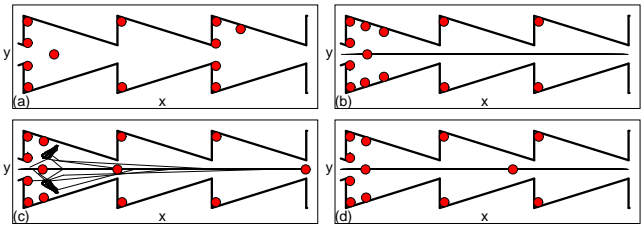}
\caption{
  The particle locations (circles) and trajectories (lines)
  for driving in the
  hard direction for the system in Fig.~\ref{fig:3}
  at $N_{c} = 3.5$.
  (a) At $F_{D} = 0.15$, the system is in a
  clogged state with a buildup of particles in the leftmost 
  plaquette.
  (b) The flow along the first velocity step at $F_{D} = 0.2$.
  (c) Multiple rearrangements of the particles occur at the
  second step in $|\langle V\rangle |$
  centered at $F_{D} = 0.26$.
  (d) The flow above the second 
  step in $|\langle V \rangle |$ at  
$F_{D} = 0.3$ where the partial clog is smaller.   
}
\label{fig:6}
\end{figure}

For $N_{c} = 3.5$,
the reentrant pinning is lost and $F_{c1}$
increases to $F_{c1}=0.175$ while the depinning 
occurs in a series of steps.
The initial flow phase disappears due to
the development of an instability within the 
pinned phase that causes particles to shift from the slanted walls to
positions near the funnel aperture prior to the onset of 1D
channel flow, 
producing a direct transition from the pinned
state to a clogged state 
without an intermediate flowing phase.
In Fig.~\ref{fig:6}(a) at $F_{D} = 0.15$,
the system is already in a clogged state even though no steady state
particle flow has occurred.
The number of particles in each plaquette varies from 2 to 5.
Just above $F_{c}$, a rearrangement of the clogged state occurs into
a partially clogged state where
a small number of particles are able to
push their way through the clog and begin to flow, as illustrated
in Fig.~\ref{fig:6}(b) for $F_{D} = 0.2$.
As the drive increases, the partially clogged configuration begins to
break apart, as shown in 
Fig.~\ref{fig:6}(c) for $F_D=0.26$ at the second jump in $|\langle V\rangle |$.
Above this jump, 
the number of particles involved in the clog
is reduced, as shown in Fig.~\ref{fig:5}(d) for $F_D=0.3$.
The size of the clog continues to decrease with further increases
in $F_D$ 
as additional particles join the flow,
leading to additional steps in $|\langle V \rangle |$.

\begin{figure}
\includegraphics[width=0.48\textwidth]{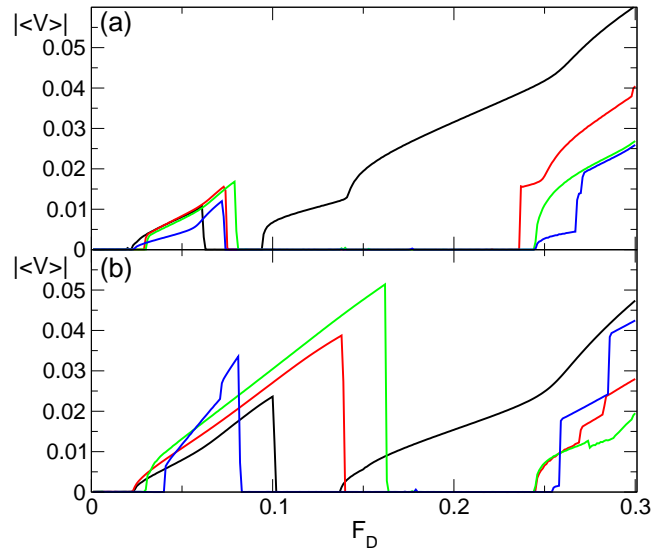}
\caption{ (a) $|\langle V\rangle|$ vs $F_{D}$
  for driving in the hard direction at
  $N_c = 3.25$ (black), 4.25 (red), 4.75 (green),
  and $5.25$ (blue).  Each curve contains
  a reentrant pinning phase.
  (b) $|\langle V\rangle |$ vs $F_D$ for $N_{c} = 5.5$ (black),
  5.75 (red), 6.0 (green), and $8.0$ (blue), also showing
reentrant pinning.  
}
\label{fig:7}
\end{figure}

In Fig.~\ref{fig:7}(a) we plot $|\langle V\rangle|$ versus
$F_{D}$  for representative fillings of 
$N_{c} = 3.25$, $4.25$, $4.75$, and $5.25$,
where in each case
there is an
interval of $F_{D}$ over which
reentrant pinning occurs.
The clogging effects become stronger
at the higher fillings,
so that
$F_{c2} \approx 0.25$.
Figure~\ref{fig:7}(b) shows $|\langle V\rangle|$ versus $F_D$
for $N_{c} = 5.5$, 5.75, 6.0, and $8.0$, which again 
exhibit reentrant pinning effects.
The onset $F_{rp}$ of the repinned state shows some variations but
generally shifts to higher
values of $F_{D}$ for $N_{c} = 5.75$ and $6.0$.

\begin{figure}
\includegraphics[width=0.48\textwidth]{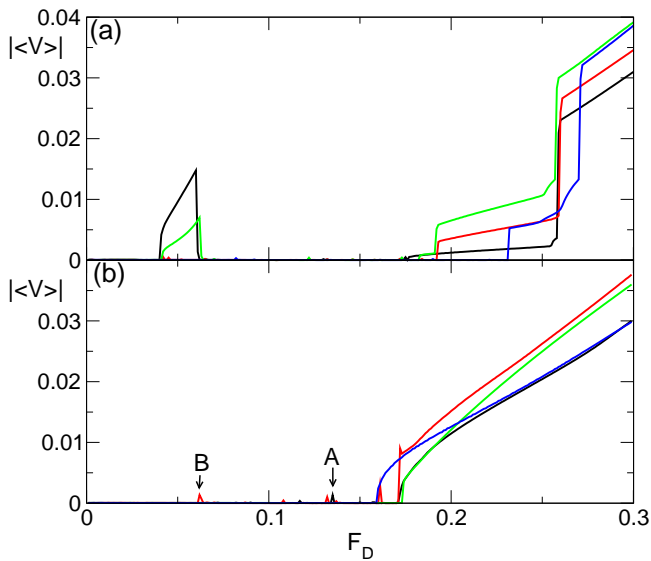}
\caption{
  $|\langle V\rangle|$ vs $F_{D}$ for driving in the hard direction at
  $N_{c} = 8.25$ (black), 8.5 (red), 8.75 (green), and $10.5$ (blue).
There is a single depinning transition for $N_{c} = 8.5$ and $N_c=10.5$,
and at these fillings
there is a direct transition from a pinned to
a clogged state without an intermediate flowing 
state. 
(b) $|\langle V\rangle|$ vs $F_D$ for $N_{c}  = 9.25$ (black), $10.75$ (red),
$11.0$ (green), and $12.0$ (blue).
The point marked A indicates the value of $F_{D}$
at which there is a transition from the pinned to the clogged state
for the $N_{c} = 9.25$ system, and
the point marked $B$ indicates the same transition for the
$N_{c} = 10.75$ system.
}
\label{fig:8}
\end{figure}

In Fig.~\ref{fig:8}(a) we plot
$|\langle V\rangle|$ versus $F_{D}$ for $N_{c} = 8.25$, 8.5, 8.75, and $10.5$.
Reentrant pinning occurs for $N_{c} = 8.25$ and $N_c=8.75$,
while at $N_{c} = 8.5$ and $N_c=10.5$,
there is only a single depinning transition
but there is a transition
from a pinned uniform state directly to
a clogged inhomogeneous state.
At  higher values of $N_{c}$
there are generally more steps
and jumps in the sliding phases
due to the increased number of particles participating in the partially clogged
arrangements, which generates 
a larger number of partially clogged stages of flow.

In Fig.~\ref{fig:8}(b) we show $|\langle V\rangle|$ versus $F_{D}$
for $N_{c} = 9.25$, $10.75$, $11.75$,
and $12.0$.
Each of these fillings exhibits only a single depinning threshold,
but the transition
from the uniform pinned state
to the inhomogeneous clogged state can still be detected in the form of
small jumps in
$|\langle V\rangle|$ in the $N_{c}= 9.25$ and
$N_c=10.75$ curves.  These jumps, marked by the letters $A$ and $B$,
are generated by a transient rearrangement of the particles in the     
pinned phase to produce a clogged structure.
For $N_{c} \geq 11$ the system transitions directly from a pinned
state to a flowing state without 
any kind of clogging.  The particle density remains uniform and
there are no jumps in $|\langle V\rangle|$, as shown in the
$N_{c} = 11.0$ and $N_c=12.0$ curves. 

\begin{figure}
\includegraphics[width=0.48\textwidth]{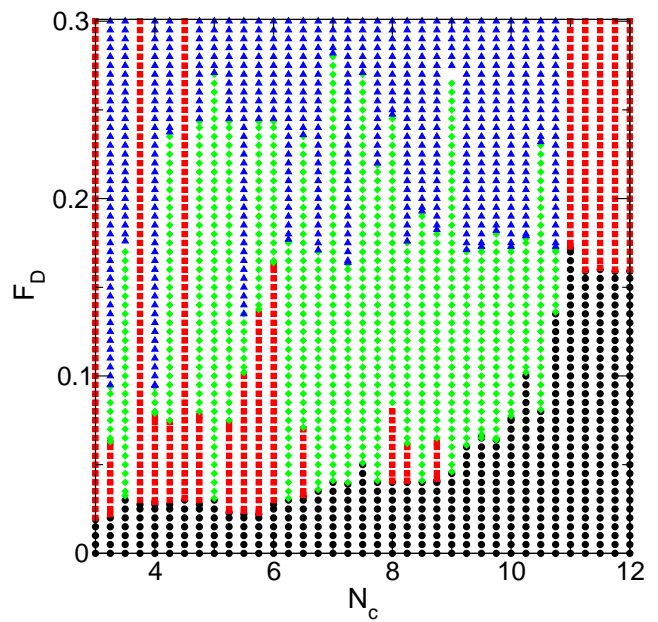}
\caption{
  Dynamic phase diagram as a function of $F_D$ vs $N_c$.
  In the pinned phase (black circles), 
  $|\langle V\rangle | = 0$ and the particle density is uniform.  
  In the sliding phases,
  $|\langle V\rangle| > 0.$  Red squares indicate the initial sliding phase
  and blue triangles indicate a reentrant sliding phase.
In the clogged or reentrant pinned phase (green diamonds),
$|\langle V\rangle| = 0$ and the particle density is heterogeneous.    
 }
\label{fig:9}
\end{figure}

In Fig.~\ref{fig:9} we plot
a dynamic phase diagram as a function of $F_D$ versus $N_c$
created from a series of simulations that highlights the pinned phase,
sliding phase, clogged phase, and reentrant sliding phase.
In the pinned state,
$|\langle V\rangle| = 0$ and the
particles are distributed evenly throughout the system.
In general, the depinning force $F_{c}$ marking the end of the pinned phase
shifts to higher $F_D$
with increasing $N_{c}$.
In the sliding states
$|\langle V\rangle| > 0$,
while the clogged state has
$|\langle V\rangle|  = 0$ with a heterogeneous distribution of particles
in the funnel array.
It is possible to identify additional dynamic regimes within the sliding state.
For example, the moving particles are evenly spaced in the sliding states at
low drives below $F_{rp}$, 
while in the sliding states for $F_{D} > F_{c2}$ 
the spacing between the moving particles is initially heterogeneous,
but gradually become more homogeneous with increasing $F_D$
as the clogs break apart.
For $N_c > 6$, the pinned state can transition directly into a clogged state with no
intermediate flowing phase, while
for $N_{c} \geq 11$, the system passes directly from a uniform pinned state to 
a uniform moving state and the clogging behavior is lost.
At low fillings $2.0 < N_{c} < 3.25$,
there is only a pinned phase and a flowing phase,
while for $3.25 < N_{c} < 11.0$
there are extended regions where clogging occurs.  We note
that due to the stochastic nature of the clogging,
the value of $F_D$ at which clogging first appears can vary
from realization to realization
when the filling is high enough for multiple particles to accumulate in
a small number of plaquettes.

\begin{figure}
 \includegraphics[width=0.48\textwidth]{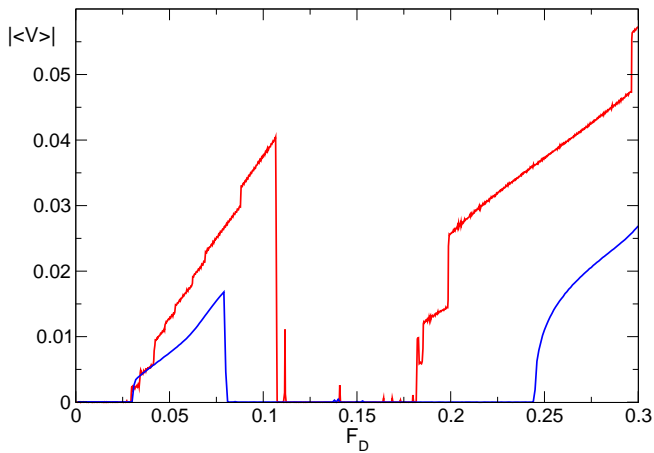}
 \caption{ $|\langle V\rangle|$ vs $F_{D}$ at $N_{c} = 4.75$ for
   driving in the hard direction at zero temperature (blue) and at
   a finite temperature of
   $F^{T} = 0.1$ (red).
    At this filling there is an extended reentrant 
    pinned region for $F^{T} = 0$.
    At finite temperature, the clogged region persists but is narrower, and
    it also contains
    a series of spikes in $|\langle V\rangle|$
    that are produced by the thermal disintegration
    of a clog which then rapidly reforms.
}
\label{fig:10}
\end{figure}

One question is how robust the clogged states
are to the introduction of a finite temperature, which could break apart
the clogged structures.
To address this we measure the velocity-force curves
at a finite temperature of $F^{T} = 0.1$,
as shown in Fig.~10 for $N_{c} = 4.75$.
At zero temperature, this filling exhibits an
extended clogged region.
Although the depinning transition at $F_{c1}$ is unaffected by the addition of
temperature, the onset of the clogged state shifts to higher
$F_D$ and the reentrant depinning
transition $F_{c2}$ shifts to lower $F_D$, giving a narrower clogged region.
Thermal fluctuations produce numerous jumps in $|\langle V\rangle|$ as
well as spikes of temporary flow within the clogged region due to
the thermally assisted disintegration of clogged structures.  Within the
reentrant pinned region, the clogs rapidly reform, but in the flowing
regimes, the clogs are permanently destroyed and the net transport through
the system is larger than at zero temperature.
In general we find
that in the range of $F_{D}$ where the clogging susceptibility is high,
clogging behavior still occurs for finite temperature
but the clogging is interspersed with jumps or avalanche events, leading
to a finite long-time average value of $|\langle V\rangle|$ which remains
much lower than the average velocity in the flowing states.

We next discuss how the clogging we consider could be observed
experimentally.
One candidate realization is superconducting vortices in nanostructured
asymmetric funnel arrays \cite{25}.
A possible issue with this system is that
the vortices are not true point particles, so
that in the clogged state
the vortices could merge 
and form multiquantum flux configurations, causing
the particle picture to break down.
Such a state could still exhibit clogging,
and there is the interesting possibility
of studying a transition
from a multiquantum clogged vortex configuration
to a flowing state, where the multiquantum vortices
would dissociate back into individual moving flux quanta.
Random point pinning would likely add a stochastic
element to the behavior of the superconducting vortices;
however, there should sill be extended regions of clogging dynamics 
at intermediate fillings.
For systems such as magnetic skyrmions or bubbles,
there is a similar issue in which the particle model can break down due to
strong spatial distortions of the skyrmion or bubble.  Such distortions could
act to prevent clogging from occurring.
Systems such as charged colloids or classical Wigner or ion crystals
are also good candidates for observing the effects we describe.
In granular or hard particle systems,
reentrant clogging could occur; however, due to the 
strong contact interactions
the system may remain in a reentrant pinned or
clogged state even for very high values of $F_{D}$,
so that it could be difficult to observe
the reentrant flowing state above $F_{c2}$.

{\it Summary --}
In summary, we have investigated a system of repulsively interacting particles
driven in the hard direction of an asymmetric funnel
array for varied ratios of the number of particles
to the number of funnel plaquettes.
Previous work on this model focused on driving in the easy 
flow direction and showed only a single depinning threshold.
In contrast, for driving in the hard direction we find
that the system can exhibit a rich variety of dynamical phases
including a reentrant pinned phase, where the system flows
at lower drives but becomes pinned when the drive is increased.
This reentrant pinning is the result of a clogging instability in which
a buildup of particles in a small number of 
plaquettes block the flow of other particles.
At higher drives, particles can force their way through the clog, leading to 
a second depinning transition.
The clogging is a robust effect that occurs over the range of
three to nearly eleven particles per plaquette. 
In addition to the reentrant pinning,
we find that in some cases a partial clogging can
produce negative differential conductivity,
and we observe
step-like features in the velocity-force curves due to the
sequential breakup of a clog.
We map the dynamic phases as a function of drive and filling fraction
and identify a pinned phase where the particles are uniformly dense and
do not move, 
a clogged phase where the particles do not move but are heterogeneously
distributed with
an accumulation of particles in certain plaquettes,
and a moving phase.
When a finite temperature is introduced, the clogging regime remains
robust but
exhibits intermittent bursts of flow when the clogs disintegrate under
thermal fluctuations
but then rapidly reform.
Finally, we describe systems in which this clogging behavior could be observed,
including superconducting vortices, colloids, Wigner crystals,
or skyrmions in asymmetric channels.  

This work was carried out under the auspices of the 
NNSA of the 
U.S. DoE
at 
LANL
under Contract No.
DE-AC52-06NA25396.

\end{document}